\documentclass[fleqn,usenatbib]{mnras}

\usepackage{newtxtext,newtxmath}
\usepackage[utf8]{inputenc}
\usepackage[T1]{fontenc}
\usepackage{graphicx}
\usepackage{amsmath}
\usepackage{booktabs}
\usepackage{upgreek}
\usepackage{mathdots}
\usepackage{mathrsfs}
\usepackage[normalem]{ulem}
\usepackage{xcolor}

\usepackage{amsmath}

\graphicspath{{./}{Figures/}}

\title[$1/f$ spectrum in MHD turbulence]{Study of the $1/f$ spectrum using virtual spacecraft sampling in MHD turbulence}

\author[M. Brodiano et al.]{
M. Brodiano,$^{1,2}$\thanks{E-mail: mbrodiano@df.uba.ar}
F. Pugliese,$^{1}$
N. Andr\'es$^{1,2}$
and P. Dmitruk$^{1,2}$
\\
$^{1}$Universidad de Buenos Aires, Facultad de Ciencias Exactas y Naturales, Departamento de F\'isica, Ciudad Universitaria, 1428 Buenos Aires, Argentina\\
$^{2}$CONICET - Universidad de Buenos Aires, Instituto de F\'isica Interdisciplinaria y Aplicada (INFINA), Ciudad Universitaria, 1428 Buenos Aires, Argentina
}

\pubyear{2026}

\begin{document}
\label{firstpage}
\pagerange{\pageref{firstpage}--\pageref{lastpage}}
\maketitle

\begin{abstract}
We investigate the appearance of a low-frequency $1/f$ magnetic spectrum in three-dimensional incompressible magnetohydrodynamic turbulence using direct numerical simulations and virtual spacecraft sampling. Our goal is to determine how the measured temporal spectra depend on the mean magnetic guide field, the probe velocity relative to the Alfv\'en speed, and the sampling angle with respect to the guide field. We find that the clearest $1/f$ ranges are obtained for stronger guide fields and are favored by faster sampler trajectories oriented more nearly perpendicular to the mean magnetic field. To characterize this behavior, we introduce a quantitative score that measures the quality and spectral coverage of the detected $1/f$ interval. We further show that, as the probe speed increases, the measured temporal spectra become progressively more consistent with a direct mapping between spatial and temporal fluctuations, particularly for nearly perpendicular sampling in the strong guide field case. These results indicate that the presence and clarity of a temporal $1/f$ range depend not only on the underlying turbulent dynamics, but also on the geometry and speed of the sampling process, with implications for the interpretation of low-frequency in situ measurements in the solar wind.
\end{abstract}

\begin{keywords}
solar wind -- magnetohydrodynamics (MHD) -- turbulence -- methods: numerical -- plasmas
\end{keywords}

\section{Introduction}

Magnetohydrodynamic (MHD) turbulence is a nonlinear multiscale system in which fluctuations evolve through the interplay of spatial structure, characteristic times and wave propagation effects (when a mean magnetic field is present) \citep{ZhouEA2004}. In such systems, the appearance of a low-frequency $1/f$ spectrum is not trivial: unlike the inertial-range dynamics, which is commonly associated with nonlinear transfer across scales, a $1/f$ range indicates the presence of correlations extending over times much longer than the characteristic turnover times of the turbulent fluctuations. Understanding how such long-period variability arises is therefore a relevant problem for MHD turbulence itself, beyond its observational manifestation in space plasmas. In the solar wind, magnetic field fluctuations display a well-known double power-law structure, with an inertial range spectrum close to \(f^{-5/3}\) at intermediate frequencies and a shallower large scale component close to \(f^{-1}\) at lower frequencies \citep{Ba1982,Burlaga_1984,BC2013,Huang2023,Davis2023,D2024}. The radial evolution of the break between these two ranges further suggests that the large scale spectrum is dynamically significant and closely connected to the origin, transport, and nonlinear dynamics of solar wind fluctuations.

The physical origin of the $1/f$ range is still under debate. One class of interpretations associates it with processes rooted in the solar atmosphere or below, such as the superposition of structures with a broad distribution of correlation scales, large scale magnetic reorganization, or signatures linked to solar rotation and coronal dynamics \citep{M1986,Matthaeus_2007,Wang2026a,Wang2026b,Wicks2013,Huang2025,Pradata_2025}. Another class of interpretations emphasizes local or in-transit mechanisms, including Alfv\'enic reflection in the expanding wind, the nonlinear interactions between outward and reflected fluctuations, and the progressive development of correlations during solar wind evolution \citep{Velli1989,Verdini2012}. Additional ideas invoke the role of parametric decay and inverse transfer, through which the nonlinear evolution of an initially higher-frequency Alfv\'enic spectrum can generate an $1/f$ range at lower frequencies \citep{C2018}. Other proposed mechanisms highlight the possible role of transverse inhomogeneity and linear wave evolution, including phase mixing and resonant absorption, in shaping large scale $1/f$ spectra \citep{Magyar_2022}. The saturation of fluctuation amplitudes under low magnetic compressibility can also naturally lead to a spectral flattening at large scales \citep{Matteini2018,Bro2026}. Altogether, current observational and theoretical results suggest that the $1/f$ range may not arise from a single universal mechanism, but rather from the combined action of source-region processes and nonlinear evolution during solar wind expansion.

Numerical studies have played an important role in clarifying which ingredients are capable of producing long-period fluctuations. In homogeneous turbulence, long-duration simulations have shown that low-frequency \(1/f\) signatures can arise in MHD systems, particularly when large scale modes contribute significantly to the dynamics and when a strong mean magnetic field is present, whereas in hydrodynamic (HD) turbulence the evidence is weaker and becomes clearer mainly in two-dimensional configurations \citep{D2007}. In the solar wind context, reduced models including reflection and nonlinear coupling in the sub-Alfv\'enic corona and solar wind have reproduced double power-law spectra and suggested that part of the $1/f$ range may be formed in these regions and then advected beyond the Alfv\'enic critical point without substantial evolution \citep{Verdini2012}. Likewise, weak-turbulence treatments of parametric instability have shown that inverse transfer toward lower frequencies can generate an $1/f$ spectrum under suitable conditions \citep{C2018}. These results demonstrate that MHD dynamics can generate large scale spectral flattening under a variety of physical conditions.

These results are particularly relevant in the context of in situ observations, because spacecraft do not measure spatial fields directly; instead, they record time series along their trajectories through a flowing plasma (e.g., \cite{Bro2023}). For this reason, synthetic spacecraft or virtual probe techniques provide a natural bridge between numerical simulations and in situ observations. By sampling the fields along prescribed trajectories, one can assess how the measured temporal spectra depend not only on the underlying turbulence, but also on the probe velocity, the angle between the trajectory and the mean magnetic field, and the degree to which a Taylor-like mapping between space and time is valid. Using synthetic measurements, \citet{Klein_2014} have shown that when the sampling speed is not sufficiently large compared with characteristic propagation speeds, intrinsic plasma frame dynamics can modify the measured frequency spectrum and limit the applicability of the frozen-flow approximation. In this sense, the observed temporal spectrum is shaped both by the plasma dynamics and by the geometry and speed of the sampling procedure itself.

In this work, we investigate the emergence of the $1/f$ range using direct numerical simulations (DNSs) of three-dimensional (3D) incompressible magnetohydrodynamic (IMHD) turbulence combined with virtual spacecraft sampling. We focus on how the measured temporal magnetic energy spectrum depends on three key parameters: the strength of the mean magnetic guide field $B_0$, the probe velocity relative to the Alfv\'en speed $V/V_A$, and the sampling angle $\theta$ with respect to the guide field. By constructing synthetic time series from a large ensemble of probes moving through statistically stationary turbulence, we seek to determine under which conditions a clear $1/f$ interval appears, how its extent varies across sampling configurations, and to what extent changes in the measured spectrum may reflect the sampling process itself. This approach allows us to isolate, within a controlled numerical framework, the roles of anisotropy, advection, and probe motion in shaping the low-frequency spectrum.

The paper is organized as follows. In Section 2, we present the dimensionless IMHD equations. In Section 3, we describe the numerical setup and the implementation of the virtual probes. In Section 4, we present the temporal spectra measured by the probes and analyze how the appearance of the $1/f$ range depends on the $B_0$, $V/V_A$, and $\theta$. We then introduce a quantitative indicator to characterize the quality and extent of the detected $1/f$ intervals and use it to summarize the trends across parameter space. Finally, we examine how the spectra measured at different probe velocities are related through rescaling arguments, and in Section 5 we summarize our main numerical results and discuss their implications for the analysis and interpretation of low-frequency spacecraft measurements in space plasmas.

\section{Incompressible MHD equations}
The 3D IMHD equations in dimensionless form are given by the momentum equation, the induction equation for the magnetic field and the solenoidal condition for both fields as, 

\begin{equation}
     \frac{\partial \mathbf{u}}{\partial t}
     + \mathbf{u}\cdot\boldsymbol{\nabla}\mathbf{u}
     = -\boldsymbol{\nabla} p
     + \mathbf{J}\times\mathbf{B}
     + \nu' \nabla^2 \mathbf{u},
     \label{NS_inc}
\end{equation}

\begin{equation}
    \frac{\partial \mathbf{B}}{\partial t}
    = \boldsymbol{\nabla}\times(\mathbf{u}\times \mathbf{B})
    + \eta' \nabla^2 \mathbf{B},
    \label{induccion}
\end{equation}

\begin{equation}
    \boldsymbol{\nabla}\cdot \mathbf{u}=0,
    \label{Solenoidu}
\end{equation}

\begin{equation}
    \boldsymbol{\nabla}\cdot\mathbf{B}=0,
    \label{SolenoidB}
\end{equation}

Here $\mathbf{u}$ is the fluctuating velocity field and $\mathbf{B}=\mathbf{B}_0+\mathbf{b}$ is the total magnetic field, where $\mathbf{B}_0$ is a uniform mean magnetic field and $\mathbf{b}$ represents the fluctuating component. The scalar field $p$ is the total pressure (including both HD and magnetic contributions), and $\mathbf{J}=\boldsymbol\nabla\times\mathbf{B}$ is the electric current density. These equations are written in dimensionless form using a characteristic length scale $L_0$, a mean scalar density $\rho_0$ and pressure $P_0$, and a typical magnetic and velocity field magnitude $b_{rms}$ and $u_{rms}=b_{rms}/\sqrt{4\pi\rho_0}$ (i.e., the r.m.s.~Alfv\'en velocity), respectively. The unit of time is defined as $t_0=L_0/u_{rms}$, corresponding to the large scale turnover time (or equivalently the Alfv\'en crossing time when $u_{rms}\sim b_{rms}$). The parameters $\nu'$ and $\eta'$ are the dimensionless viscosity and magnetic diffusivity, respectively, which can be expressed in terms of the kinetic and magnetic Reynolds numbers as $\nu'=1/R_e$ and $\eta'=1/R_m$, where $R_e=u_{rms}L_0/\nu$ and $R_m=u_{rms}L_0/\eta$.

\section{Numerical setup and virtual spacecraft sampling} 

To investigate the emergence of the $1/f$ range, we performed direct numerical simulations (DNSs) of the IMHD equations in the presence of a uniform magnetic guide field $\mathbf{B}_0=B_0~{\hat z}$ in a three-dimensional periodic domain of size $2\pi$, using spatial resolutions of $128^3$ and $256^3$ grid points (here, we show only results from the $256^3$ simulations). The IMHD Eqs.~\eqref{NS_inc}-\eqref{SolenoidB} were solved with the Fourier pseudospectral code \textsc{GHOST} \citep{Go2005b,Mi2011}. Time integration was carried out with a second order Runge-Kutta scheme, which ensures exact energy conservation for the continuous time, spatially discrete equations \citep{Mi2011}. Each simulation was evolved for a total duration of $800\,t_0$. This long integration time is necessary to properly resolve the low-frequency part of the magnetic energy spectrum and to access the large scale regime characterized by a power-law behavior close to $E(f)\propto f^{-1}$.

For simplicity, we considered identical dimensionless viscosity and magnetic diffusivity, $\nu'=\eta'=1.7\times10^{-3}$, corresponding to a magnetic Prandtl number $P_m=1$, typical for the solar wind \citep{Osman2011,Pecora_2023,Pugliese2023,Bro2021}. To reach a statistically stationary turbulent state, both the velocity field and the magnetic vector potential were externally forced. A mechanical forcing $\mathbf{F}$ was applied to the velocity field, while an electromotive forcing $\boldsymbol{\epsilon}$ was applied to the magnetic vector potential. These forcings are uncorrelated and inject neither kinetic nor magnetic helicity. At $t=0$, for each forcing function, a random three-dimensional isotropic field $f_{\mathbf{k}}$ is generated in Fourier space by populating all modes within a spherical shell $2<k<3$ with fixed amplitude $f$ and assigning a random phase $\phi_{\mathbf{k}}$ to each wavevector $\mathbf{k}$. This forcing procedure continuously injects energy at the largest scales of the system.

\begin{figure}
    \centering
    \includegraphics[width=\columnwidth]{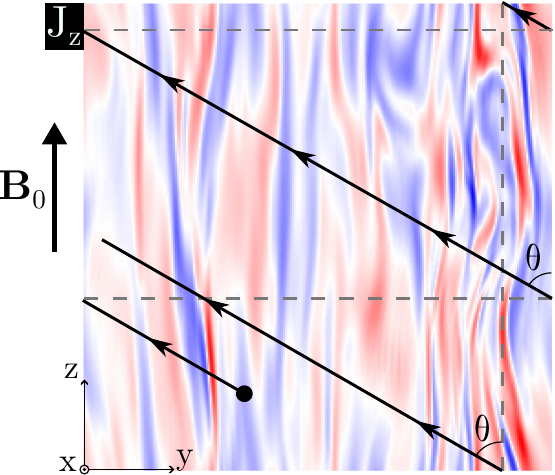}
    \caption{Schematic representation of the virtual probe sampling procedure. Virtual probes move through the periodic simulation domain along straight trajectories with prescribed constant velocities, forming an angle $\theta$ with respect to $\mathbf{B}_0$. Along each path, the probes sample the turbulent fields at fixed time intervals, generating synthetic time series analogous to in situ spacecraft measurements. The background colorbar represents $z$ component of the current density.
    }
    \label{fig:probe_scheme}
\end{figure}

To simulate \textit{in situ} measurements in the solar wind, we implemented a set of virtual probes within the simulation domain. These probes move through the computational box with prescribed constant velocities, sampling the plasma fields along their trajectories. 
In particular, we considered a total of 440,000 probes with different initial velocities and different orientations with respect to the guide field $\mathbf{B}_0$, with angles $\theta$ spanning the range from $0^\circ$ to $90^\circ$. For each combination of probe speed and sampling angle $\theta$, the probes were uniformly distributed in their initial positions within the computational domain and in the azimuthal angle with respect to the guide field direction. Specifically, the initial velocity was varied every 40,000 probes, and for each fixed velocity the sampling angle was assigned every 4,000 probes. In this way, we considered 11 different velocities and 10 different angles for each velocity. The sampling geometry is illustrated schematically in Figure~\ref{fig:probe_scheme}. For both the weak guide field case ($B_0=1$) and the strong guide field case ($B_0=6$), the measured field time series were sampled with the same cadence, $\Delta t = 2.5\times10^{-3}$. We refer to the weak magnetic field case as Run I and to the strong case as Run II.
The choice of probe velocities is best understood in terms of the ratio $V/V_A$, where $V$ is the probe speed and $V_A$ is the Alfv\'en speed associated with each simulation. Because $V_A$ increases with the guide field intensity, the same sampling procedure and comparable absolute probe velocities correspond to markedly different values of $V/V_A$ in the two runs. As a result, in Run I case we were able to explore a much broader range, reaching $V/V_A \sim 10$, whereas in Run II case the explored range extends only up to $V/V_A \sim 3$, which is consistent with the spacecraft speed relative to the solar wind speed (typically, $V \sim 400 \ \rm{km/s}$ and $V_A \sim 150 \ \rm{km/s}$). This distinction is important because the appearance and quality of the $1/f$ range depend not only on the sampling angle but also on the ratio between the probe velocity and the characteristic Alfv\'enic propagation speed. In this way, the selected set of probe velocities allows us to assess how the measured spectra change across different sampling regimes, from relatively slow to effectively super-Alfv\'enic trajectories.

By varying both the probe speed and its angle $\theta$ relative to $\mathbf{B}_0$, we explored a broad set of sampling configurations and analyzed under which conditions a $1/f$ magnetic energy spectrum emerges. This procedure enables the construction of synthetic spacecraft time series directly from the numerical simulations and provides a framework for diagnosing turbulent fluctuations using methods commonly applied to observational data.

\section{Results and discussion} 

\subsection{The role of the magnetic guide field in the presence of $1/f$ spectrum}

\begin{figure}
    \centering
    \includegraphics[width=\columnwidth]{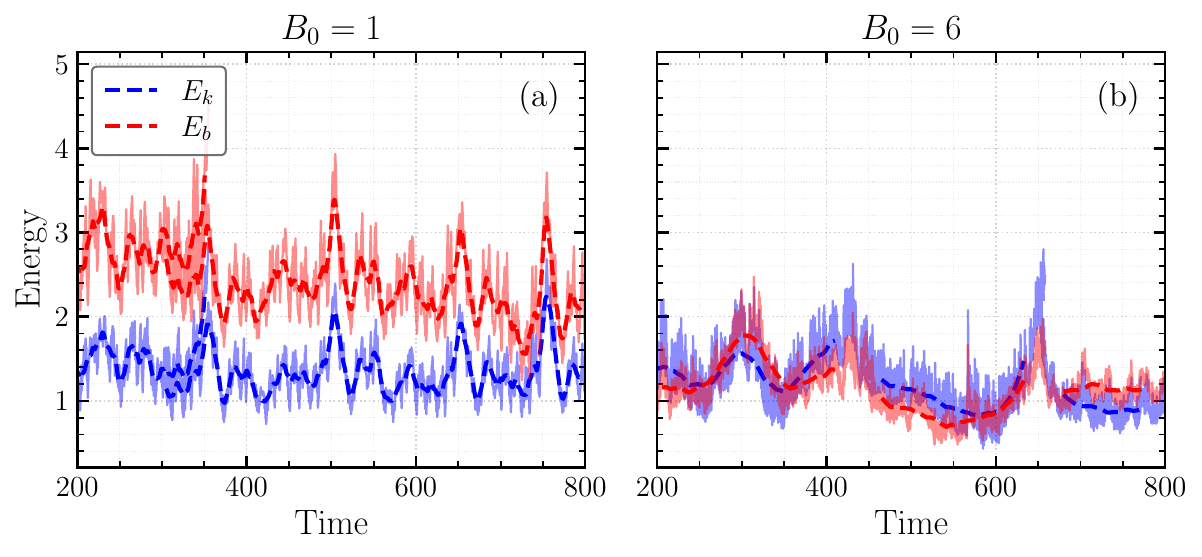}
    \caption{Temporal evolution of the kinetic and magnetic energies, $E_k(t)$ and $E_b(t)$, for (a) Run I ($B_0=1$) and (b) Run II ($B_0=6$), after discarding the initial transient. Solid lines show the instantaneous energy time series, while dashed lines correspond to running averages computed over a fixed number of points and are included for visualization purposes, highlighting the large-scale temporal trends in the energy evolution.
    }
    \label{fig:globals}
\end{figure}

As previously mentioned, the simulations performed in this work were analyzed only after a statistically stationary regime had been reached, in which the total energy and the total dissipation rate of the system (not shown here) exhibit a stable temporal evolution. In particular, we monitored the full time series of the kinetic and magnetic energies, $E_k(t)$ and $E_b(t)$, for both magnetic guide field amplitudes. Figure~\ref{fig:globals} (a) and (b) show the temporal evolution of these quantities during the simulation, after discarding the initial transient. In both cases, the energies fluctuate around well-defined mean values over the entire integration interval. Figure~\ref{fig:globals} also reveals a qualitative difference between the two magnetic guide field runs. For Run I (panel (a)), both $E_k$ and $E_b$ display comparatively faster and more irregular fluctuations, with shorter-time variations superimposed on the mean level. By contrast, for Run II (panel (b)), the global energies evolve more smoothly and exhibit broader, longer-time modulations.

\begin{figure*}
    \centering
    \includegraphics[width=\textwidth]{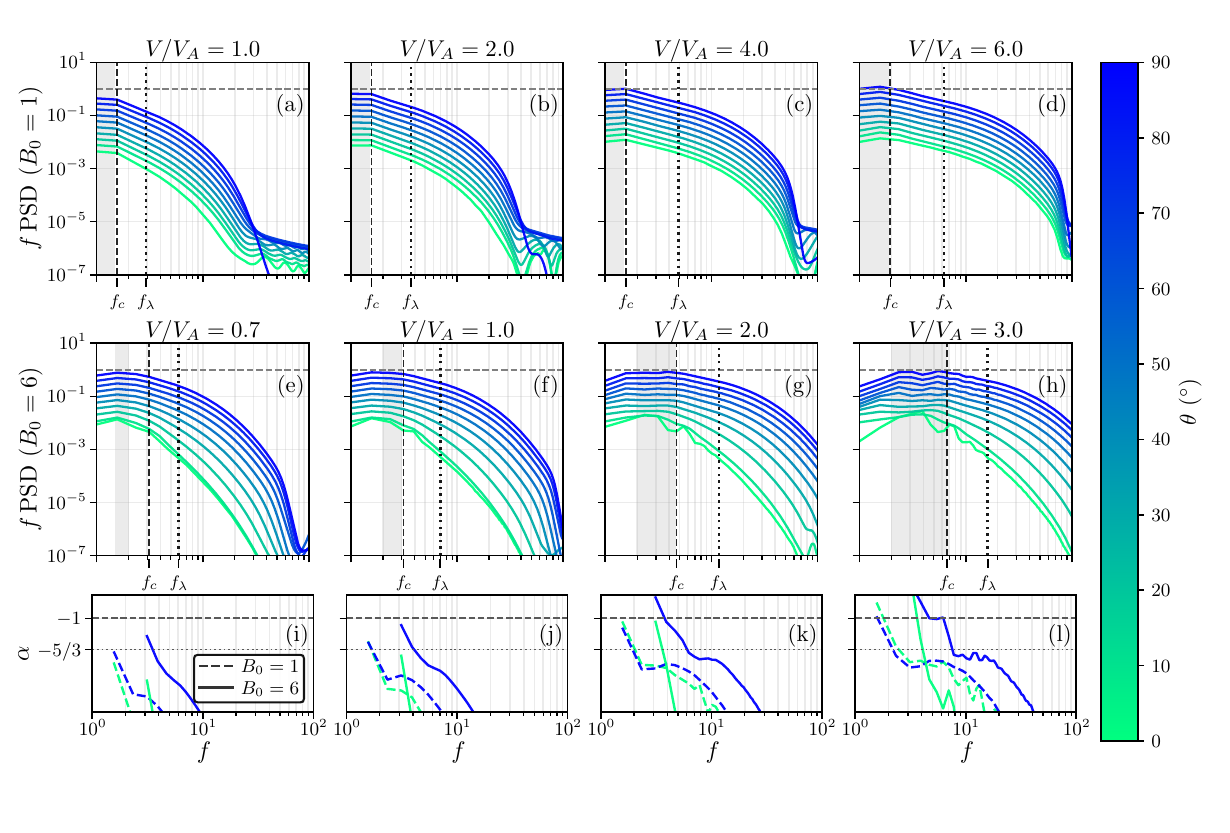}
    \caption{Compensated magnetic energy spectra measured by the virtual probes for Run I ($B_0=1$, first row) and Run II ($B_0=6$, second row). Each column corresponds to a fixed probe velocity, normalized by the Alfv\'en speed, $V/V_A$. The colored curves represent different sampling angles between the probe trajectory and the magnetic guide field, from $0^\circ$ to $90^\circ$, as indicated by the colorbar. For clarity, a vertical offset has been applied between curves. The gray shaded regions denote the frequency intervals used to identify the candidate $1/f$ ranges, while the black dashed and dotted vertical lines indicate the correlation frequency $f_c$ and the Taylor frequency $f_\lambda$, respectively. The third row shows the locally fitted spectral slope $\alpha(f)$ for $\theta=0^\circ$ and $\theta=90^\circ$, using the same colors as in the spectra. Dashed lines correspond to Run I and solid lines to Run II. The gray dashed and dotted horizontal lines indicate the reference slopes $\alpha=-1$ and $\alpha=-5/3$, respectively.
}
    \label{fig:specB01B06}
\end{figure*}

Figure \ref{fig:specB01B06} shows the average (in probes) temporal magnetic power spectral density (PSD) measured by the virtual probes for a range of angles and probe velocities, for Run I (a-d) and Run II (e-h), respectively. Each panel displays the compensated magnetic spectrum as a function of frequency for a fixed value of the probe velocity $V/V_A$, normalized by the Alfvén speed $V_A$. The different colored curves correspond to different angles $\theta$ between the probe trajectory and the mean magnetic field, ranging from $0^\circ$ (light green) to $90^\circ$ (blue). For visualization purposes, the spectra have been vertically offset, allowing the angular dependence of the spectral trends to be more clearly distinguished. The gray shaded region indicates the frequency interval considered as a candidate $1/f$ range, while the vertical dashed and dotted lines mark the correlation frequency $f_c$ and the Taylor frequency $f_\lambda$, respectively. The correlation frequency is estimated from the temporal magnetic spectra as the frequency at which the spectrum shows a transition between the $1/f$ range and the steeper inertial range (e.g., \cite{Ghanbari_2023}), while the Taylor frequency is computed (for each temporal spectrum) as $f_\lambda = \left(\int f^2 {\rm PSD}(f)\,df/\int {\rm PSD}(f)\,df \right)^{1/2}$ and the value shown corresponds to the median over sampling angles for each fixed probe velocity. This definition is the spectral counterpart of the Taylor scale estimated by \citet{Chuychai2014}. This frequency is therefore used as a temporal reference scale to indicate the transition toward smaller-scale fluctuations within the probe spectrum.

In this representation, a horizontal plateau indicates a spectral scaling close to $E(f)\propto f^{-1}$, so that the extent of the plateau within the shaded region provides a direct measure of the presence of the $1/f$ range. The comparison between panels therefore allows us to evaluate how this low-frequency regime depends on the guide field amplitude, the probe speed, and the sampling direction. In addition, in Figure \ref{fig:specB01B06} (i-l), the locally fitted spectral slope is showed as a function of frequency, which provides a complementary diagnostic of the frequency intervals where the spectra are consistent with a $1/f$ scaling or depart toward steeper slopes.

We first consider Run I, for which the spectra (a-d) exhibit only a limited low-frequency plateau and a weak dependence on the sampling direction, particularly at the lowest probe speeds. For $V/V_A=1$, the spectra corresponding to different angles remain similar over a broad frequency interval. Even at higher relative probe speeds, however, the development of an extended $1/f$ range remains limited. Although some curves at larger angles display a flatter low-frequency behavior, the plateau remains narrow and is followed by a transition toward steeper slopes. Moreover, in panels (i-l), the local slopes show only short intervals close to $\alpha=-1$, followed by a rapid departure toward more negative values, and the separation between different sampling directions remains weak. Thus, in Run I, increasing the probe speed alone is not sufficient to produce a broad $1/f$ interval across all directions. This suggests that, for a weak guide field, the low-frequency spectral organization is weaker and the $1/f$ regime is less developed. Under these conditions, the distinction between parallel and perpendicular sampling is reduced, indicating weaker anisotropy of the fluctuations.

A different behavior is found in Run II, where the dependence on the sampling direction is much clearer. For all explored probe velocities (e-h), the spectra display a systematic ordering: as the trajectories become more perpendicular to the mean magnetic field, the compensated spectra become flatter at low frequencies and the corresponding plateau extends over a wider interval. This trend is already visible at the lowest velocity shown, $V/V_A=0.7$ (panel e), and becomes progressively more pronounced as the probe speed increases to $V/V_A=1$, $2$, and $3$ (panels f-h). Thus, the extension of the candidate $1/f$ range increases not only with the degree of perpendicular sampling, but also with the relative probe velocity. The curves associated with nearly perpendicular trajectories show the widest plateaus, while more parallel samplings depart earlier from the low-frequency regime. This is further supported by the local spectral slopes (panels i-l) exhibiting an extended interval close to $\alpha=-1$ for the highest velocities and most perpendicular directions, followed by a transition to the inertial range. Therefore, in the presence of a strong guide field, the observed temporal spectrum is controlled by the combined effect of sampling geometry and probe speed, with faster perpendicular trajectories favoring the clearest development of a broad $1/f$ interval.

To complement the spectral analysis discussed above, we constructed two-dimensional maps to quantify how clearly a $1/f$ range develops as a function of probe velocity and sampling angle. Although compensated spectra provide a direct visual indication of flat low-frequency plateaus, a quantitative criterion is useful to compare all sampled configurations in a systematic way. For this purpose, we define a $1/f$ score based on the compensated temporal magnetic spectrum, $Y(f)=fE(f)$, within a prescribed candidate frequency band.

The score is obtained by performing sliding-window linear fits in log-log coordinates within this band. At each window, we estimate the local slope, the root-mean-square residual scatter, and the compensated amplitude. A window is considered compatible with a $1/f$ plateau when the local slope remains close to the expected value, the residual scatter is sufficiently small, and the amplitude stays above a fixed fraction of the maximum compensated amplitude in the candidate band. The latter condition avoids selecting weak high-frequency tails as part of the plateau. The method then identifies the longest contiguous set of valid windows and performs a global fit over the corresponding frequency interval.

The frequency extent of the detected interval is measured in logarithmic units as
\begin{equation}
    \mathrm{width}_{\rm dec}=\log_{10}\!\left(\frac{f_2}{f_1}\right),
\end{equation}
where $f_1$ and $f_2$ are the lower and upper limits of the selected range. The final score combines the relative coverage of the candidate band with penalties for deviations from the ideal $1/f$ slope and for large residual fluctuations:
\begin{equation}\label{score}
    \mathrm{score} = \mathrm{coverage}^{\,p}\times \mathrm{flatness}\times \mathrm{smoothness},
\end{equation}
with
\begin{equation}
    \mathrm{coverage}=\frac{\mathrm{width}_{\rm dec}}{\log_{10}(f_{\rm max}/f_{\rm min})}.
\end{equation}
The factor $p$ in Eq.~\eqref{score} controls how strongly broad frequency intervals are favored, and is set to $p=1.3$. Thus, the score is a composite indicator that increases when the detected range is broad, close to the expected $1/f$ scaling, and sufficiently smooth.  

\begin{figure*}
    \centering
    \includegraphics[width=\textwidth]{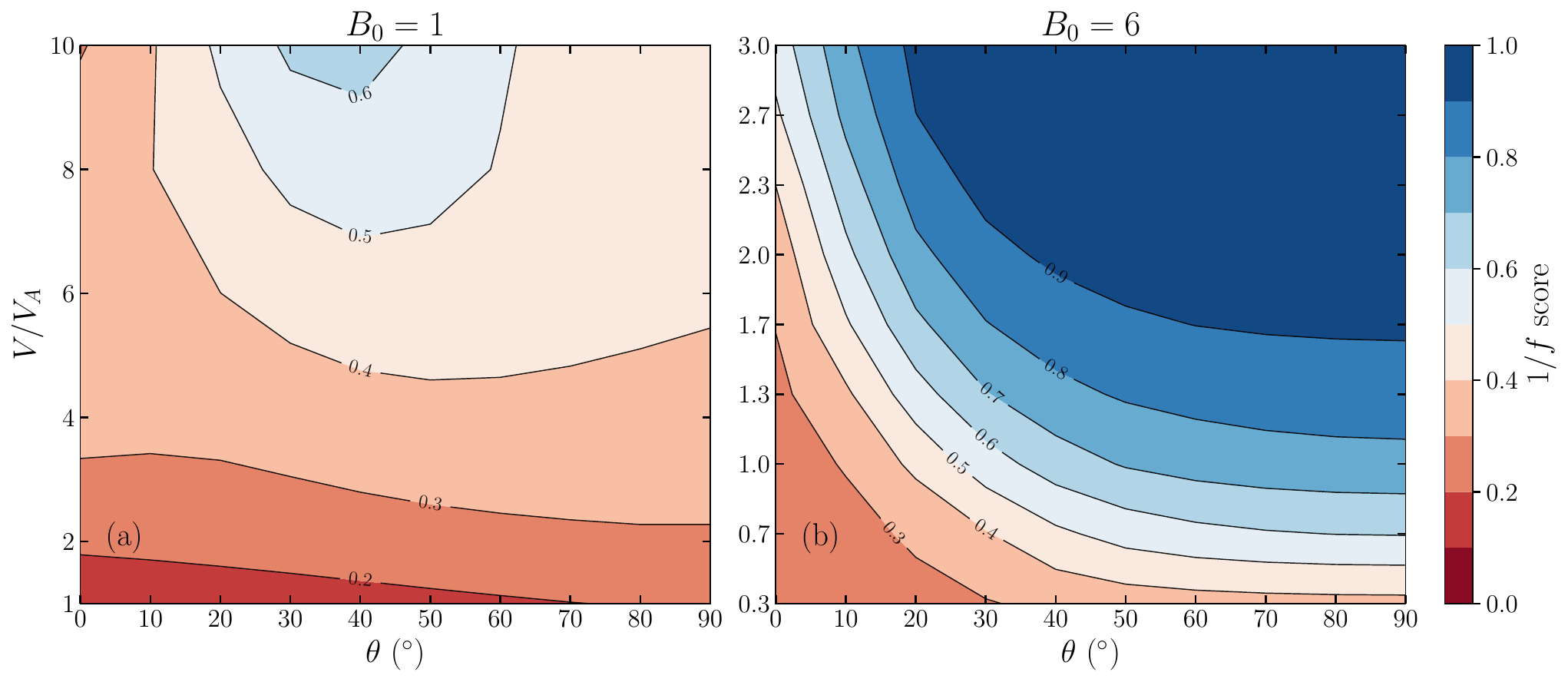}
    \caption{Two-dimensional maps of the $1/f$ score for (a) Run I ($B_0=1$) and (b) Run II ($B_0=6$), as a function of the sampling angle $\theta$ and the normalized probe velocity $V/V_A$. For visualization purposes, black contour lines indicate selected score levels.}
    \label{fig:scoreB01B06}
\end{figure*}

The resulting score maps are shown in Figure~\ref{fig:scoreB01B06} (a) and (b). To facilitate the visualization of the global trends, the discrete score values were smoothed using a two-dimensional kernel filter, and the resulting fields are displayed together with contour lines indicating selected score levels. In Run I (a), the score remains generally modest over most of the explored parameter space. The lowest values are found at the smallest probe velocities, where the spectra show only a very limited low-frequency flattening. As $V/V_A$ increases, the score also tends to increase, indicating that the detected $1/f$-like interval becomes somewhat broader and better defined. However, the angular dependence remains weak: rather than showing a systematic enhancement toward nearly perpendicular sampling, the score is relatively similar across $\theta$ for a given velocity, consistent with a more isotropic distribution of fluctuations. A different pattern is observed in Run II (b), as the score map exhibits a clear and systematic increase with both $V/V_A$ and $\theta$. The lowest scores are again associated with slow and nearly parallel sampling, but the score rises steadily as the trajectories become faster and more perpendicular to the guide field. This trend is clearly highlighted by the contour lines, which show a progressive displacement of higher score levels toward larger $V/V_A$ and $\theta$. This behavior indicates that, in the strong guide field regime, the detected $1/f$-like intervals are not only flatter, but also more extended and smoother than in Run I. In other words, the score map quantitatively confirms that a strong field, combined with sufficiently fast and nearly perpendicular sampling, produces the clearest temporal signature of the $1/f$ range.


\subsection{Probe velocity and Taylor hypothesis}

The results presented in the previous section suggest that the observed temporal spectra may be shaped, at least in part, by the way the probe samples the spatial structure of the turbulent fluctuations. We therefore examine to what extent the synthetic probe measurements are consistent with the Taylor hypothesis \citep{Taylor1938}. In few words, under the Taylor hypothesis, the temporal variations measured along the probe trajectory are assumed to be dominated by advection, so that the intrinsic temporal evolution of the turbulent structures can be neglected compared with the relative motion of the probe through the flow. In this case, frequency $f$ and wavenumber $\bf k$ are related through,
\begin{equation}
    f \sim \mathbf{k}\cdot\mathbf{V},
\end{equation}
and the temporal spectrum measured by the probe can be directly associated with a spatial spectrum along the sampling direction. This approximation is expected to perform better when the probe speed is sufficiently large compared with the characteristic propagation speed of the fluctuations. In the present problem, this condition is naturally expressed in terms of the ratio $V/V_A$, since $V_A$ sets the relevant Alfv\'enic propagation scale of the system.


In our synthetic probe measurements, this idea can be tested by rescaling the probe frequency as a wavenumber using the probe speed $V$, namely $k\sim f/V$, and comparing the resulting spectra with the corresponding Eulerian spectra. Here, the Eulerian reference corresponds to the isotropic spatial magnetic energy spectra parallel and perpendicular to the guide field, obtained from Fourier space spectral outputs at fixed times and then averaged over all available snapshots. In this way, we obtain the mean Eulerian spectra $E(k_\parallel)$ and $E(k_\perp)$, which characterize the average spatial distribution of magnetic fluctuations in the two directions. Unlike the probe spectra, which are constructed from temporal signals measured along moving trajectories, these Eulerian spectra contain only spatial information and therefore provide the natural reference for assessing the validity of the Taylor hypothesis.

\begin{figure*}
    \centering
    \includegraphics[width=\textwidth]{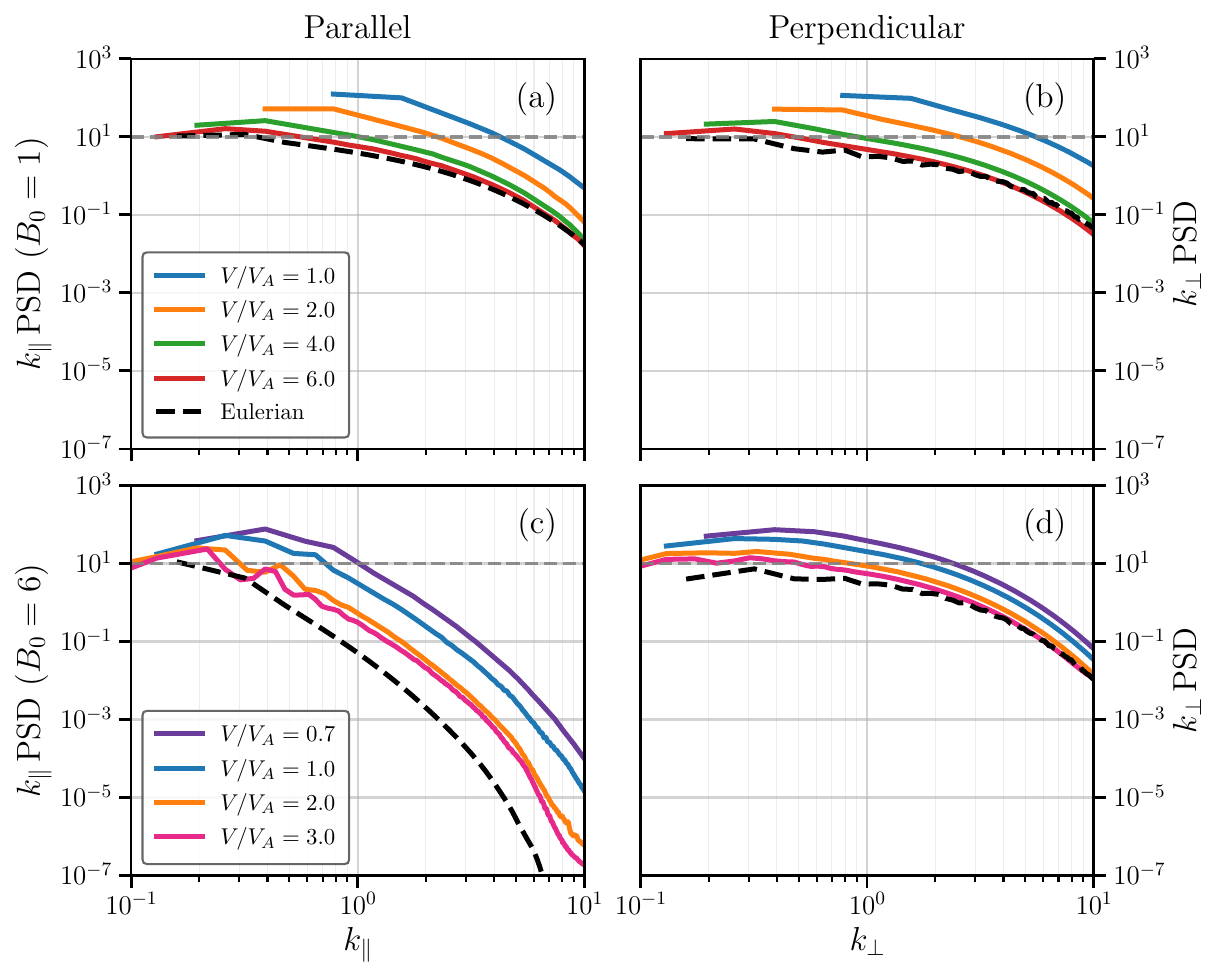}
    \caption{Comparison between temporal and spatial (Eulerian) compensated magnetic energy spectra. Panels (a) and (b) correspond to Run I ($B_0=1$), while panels (c) and (d) correspond to Run II ($B_0=6$). The left and right columns show spectra obtained from probes sampling parallel and perpendicular to the guide field, respectively. The colored curves represent the compensated temporal spectra measured at different normalized velocities $V/V_A$, after converting the temporal frequency into a spatial wavenumber using the probe speed. The dashed black curves denote the corresponding compensated Eulerian magnetic energy spectra.}
    \label{fig:taylor_B01B06}
\end{figure*}

Figure~\ref{fig:taylor_B01B06} shows the compensated magnetic energy spectra for  Run I (a and b) and Run II (c and d), separated into parallel and perpendicular sampling with respect to the guide field. The colored curves correspond to the spectra measured by virtual probes moving at different normalized velocities. Note that the wavenumber is normalized by $2\pi$ so that the Taylor-converted frequency, $k \sim f/V$, is expressed in the same units as the Fourier wavenumber used for the Eulerian spectra. The dashed black curves represent the corresponding Eulerian compensated magnetic energy spectra.
For Run I (a and b), increasing $V/V_A$ leads to a progressively better agreement between the probe spectra and the Eulerian spectra in both directions, particularly at the smallest scales. For the lowest values of $V/V_A$, the probe spectra remain systematically above the Eulerian reference over a broad range of wavenumbers, indicating that the temporal signal still contains an important contribution from the intrinsic dynamics of the fluctuations. As the velocity increases, the rescaled spectra recover more closely the overall shape of the Eulerian spectrum. The distinction between the parallel and perpendicular directions is relatively modest, consistent with the weaker anisotropy discussed in the previous section.

For Run II (c and d), a clearer directional dependence is found, particularly for perpendicular sampling. As well as the previous case, the agreement between the probe spectra and the Eulerian spectrum improves systematically as the probe velocity increases. For the largest values of $V/V_A$, the rescaled temporal spectra recover a shape that is much closer to the Eulerian reference over an extended range of wavenumbers, especially at scales associated with the inertial and dissipative ranges. This indicates that, for sufficiently fast perpendicular sampler, the temporal signal is increasingly dominated by advection and can be more reliably interpreted in spatial terms. However, this correspondence breaks down at the largest scales. In the low-wavenumber range, the temporal spectra display a clear $1/f$ contribution that is not mirrored by a corresponding $1/k$ scaling in the Eulerian spectrum. Thus, the Taylor-like conversion does not imply a one-to-one mapping between the temporal $1/f$ range and a pre-existing spatial power-law at low $k$. Rather, the sampling procedure generates temporal power in a range of effective wavenumbers where the Eulerian spectrum does not show an analogous scaling. In this sense, the low-frequency $1/f$ range appears as a feature of the long-time probe signal, shaped by the repeated sampling of the turbulent domain, rather than as the direct image of an Eulerian low-wavenumber range. By contrast, in the parallel direction, the collapse is substantially poorer. Even for the fastest probes, significant differences with respect to the Eulerian spectrum remain across the full range of scales, thereby limiting the validity of the Taylor approximation.

\subsection{Spectral rescaling under Taylor hypothesis}

The dependence of the measured temporal spectra on the probe velocity can be understood, to first approximation, as a sampling effect under the Taylor hypothesis. Consider a probe moving at constant speed $V$ through a spatial field $u(x)$, so that the recorded signal is
\begin{equation}
    s(t)=u(x=Vt).
\end{equation}

If the same field is instead sampled by a probe moving at speed $V'$, the measured time series becomes
\begin{equation}
    s'(t)=u(V't)=s\!\left(\frac{V'}{V}t\right).
\end{equation}

Thus, changing the probe speed does not modify the underlying spatial structure being sampled, but only changes the rate at which that structure is traversed in time. In this sense, the temporal signal is dilated or compressed depending on the ratio $V'/V$. This simple time rescaling has a direct spectral consequence. If two probe signals correspond to the same spatial field sampled at different velocities, their PSDs are related by the standard scaling property

\begin{equation}
    \text{PSD}_{V\rightarrow V'}(f)=\frac{V}{V'}\,\text{PSD}_V\!\left(\frac{V}{V'}f\right).
\end{equation}
Therefore, increasing the probe velocity shifts the temporal spectrum toward higher frequencies and rescales its amplitude accordingly. This relation provides a simple way to predict how the spectrum measured at one probe speed should transform when the same fluctuations are observed at another speed. In this framework, the dependence of the temporal spectrum on $V$ arises mainly from the change in sampling rate by a factor $V/V'$, rather than from a modification of the underlying spatial fluctuations. 

\begin{figure}
    \centering
    \includegraphics[width=1\linewidth]{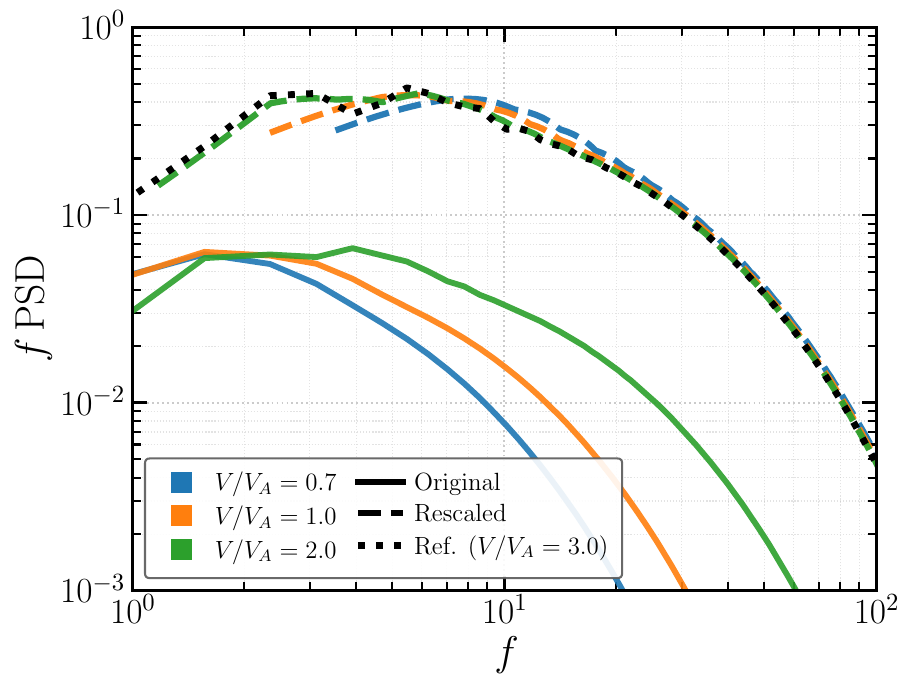}
    \caption{Test of the sampling-rescaling relation for the temporal magnetic spectra in Run II ($B_0=6$). The compensated spectrum measured at $V/V_A=3$ is used as the reference and compared with the spectra measured at lower probe velocities, $V/V_A=0.7$, $1$, and $2$, together with their rescaled counterparts. Solid lines show the original spectra, dashed lines the rescaled spectra, and the dotted black line the reference spectrum. Each color corresponds to a different probe velocity. A vertical offset has been applied to the original spectra for visualization purposes.}
    \label{fig:collapse_sampling}
\end{figure}

To test this idea, we take one measured spectrum as a reference and use the previous expression to predict its form at a different sampling velocity. The result is then compared with the spectrum obtained directly from probes moving at that target speed. Figure~\ref{fig:collapse_sampling} shows the velocity rescaling procedure for Run II. The compensated temporal spectrum measured at $V/V_A=3$ is taken as the reference, while the spectra obtained at lower probe velocities, $V/V_A=0.7$, $1$, and $2$, are shown together with their rescaled counterparts. Each color corresponds to a different probe velocity. Solid color lines indicate the original spectra, dashed color lines represent the rescaled spectra, and the dotted black line denotes the reference spectrum at $V/V_A=3$. For visualization purposes, a vertical offset has been applied to the original spectra, allowing the frequency displacement associated with the rescaling to be more clearly compared. Once the rescaling is applied, the remapped curves shift in frequency and collapse much more closely onto the reference spectrum over a broad frequency interval. The agreement is particularly good for the highest of the rescaled velocities, while larger discrepancies remain for the slowest probes. This behavior indicates that, when the probe velocity is sufficiently large, the effect of changing the sampling speed can be described to a good approximation as a dilation or compression of the temporal signal. Conversely, for slower probes, the rescaling is less successful, suggesting that the measured time series retains a stronger contribution from the intrinsic temporal dynamics of the fluctuations. This result is consistent with the previous analysis: the mapping between spatial and temporal scales becomes more reliable at larger probe velocities, while deviations from a simple Taylor-like interpretation become more important in the slow sampling regime.

Overall, this analysis shows that a substantial part of the velocity dependence of the measured temporal spectra can be interpreted as a consequence of the sampling procedure itself. In particular, the collapse obtained for the fastest probes supports the idea that, in that regime, the Taylor hypothesis provides a useful approximation for relating spectra measured at different probe velocities. At the same time, the residual differences observed at lower $V/V_A$ indicate that this approximation becomes less accurate, in agreement with the behavior discussed above.

\section{Discussion and Conclusions} 

We examined how the emergence of a low-frequency $1/f$ range depends on the amplitude of the magnetic guide field. To this end, we compared the temporal magnetic spectra measured by virtual probes in Run I ($B_0=1$) and Run II ($B_0=6$), considering different sampling velocities and angles with respect to the mean magnetic field. The compensated spectra shown in Figure~\ref{fig:specB01B06} reveal that Run I develops only limited low-frequency plateaus, with a relatively weak dependence on the sampling direction. By contrast, in Run II, the spectra exhibit a much clearer angular organization: nearly perpendicular trajectories show flatter and more extended $1/f$ range, and this behavior becomes more pronounced as the probe velocity increases. This trend is also reflected in the score maps, which show modest and weakly angle-dependent values for Run I, but a systematic increase of the $1/f$ score with both $V/V_A$ and $\theta$ for Run II. These results indicate that the presence of a clear temporal $1/f$ range is not determined by the guide field strength alone, but rather by its combined effect with the sampling geometry and probe speed.

These findings are broadly consistent with the previous numerical study by \citet{D2007}, who also found that the low-frequency $1/f$ range becomes more prominent as the mean magnetic field is increased. In that work, compensated magnetic energy spectra measured in the direction perpendicular to the guide field, using fixed probes located in a plane at the center of the simulation domain, showed that the $1/f$ interval was almost absent for small $B_0$, but became progressively clearer for larger guide field amplitudes, together with the appearance of longer-time fluctuations in the corresponding time series. A similar trend is recovered here, as broader long-time modulations are observed in the global energy time series for the strong guide field case (see Figure~\ref{fig:globals}).

The numerical trends described above can be placed in the broader context of recent \textit{in situ} observations in the inner heliosphere. Indeed, large scale magnetic spectrum is known to evolve with distance from the Sun: the break between the low-frequency $1/f$ range and the inertial interval shifts toward lower frequencies with increasing heliocentric distance, a trend usually associated with the growth of the correlation length and the broadening of the outer scale \citep{BC2013,Chen_2020}. More recent \textit{Parker Solar Probe} (PSP) studies have further shown that this evolution is better organized when intervals are ordered by the advection time, $t_{\rm adv}=R/V_{\rm sw}$, rather than by heliocentric distance alone, and that the low-frequency spectral index tends to approach values close to $-1$ during the outward evolution of the young solar wind \citep{Huang2023,Davis2023}. Although a clear $1/f$ range is most often associated with fast, highly Alfv\'enic streams, carefully conditioned slow wind intervals can also display a full $1/f$ range at 1 au \citep{Dorseth2024b}. Therefore, these results suggest that the large scale spectrum is shaped not only by the local turbulent dynamics, but also by the expansion history, the advection speed, and the way the fluctuations are sampled.

In that observational context, our results are qualitatively consistent with the observational picture that the development of the large scale spectrum depends both on the radial evolution of the solar wind and on how fluctuations are advected and sampled. Observationally, the low-frequency spectral properties evolve with heliocentric distance, while the magnetic field amplitude tends to decrease outward through the inner heliosphere \citep[see, e.g.,][]{Sa2023}. In our simulations, the clearest $1/f$ range appears for higher probe speeds and in the strong guide field case. Although the probe speed in our simulations is not directly equivalent to the solar wind bulk speed, both quantities play an analogous role in controlling how spatial structure is converted into temporal fluctuations. Likewise, the strong guide field case provides a simplified analogue of the observational tendency for the magnetic field to be larger closer to the Sun. From this perspective, the comparison suggests that, just as the low-frequency spectrum in the solar wind is shaped by outward evolution, changing magnetic conditions, and advection speed, the emergence of a $1/f$ range in the simulations is favored by a stronger mean magnetic field and faster sampler.

The dependence on probe speed discussed above also has implications for the interpretation of the measured temporal spectra. Since the clearest $1/f$ ranges are obtained under specific sampling conditions, the low-frequency temporal signal cannot be interpreted independently of how the turbulent fields are sampled. The comparison with Eulerian spectra (see Figure~\ref{fig:taylor_B01B06}) shows that the relation between temporal and spatial fluctuations is not uniform across scales. While the probe spectra become more consistent with the spatial organization of the turbulence when the sampling is perpendicular and sufficiently fast, especially at intermediate and small scales, the largest scales retain a low-frequency temporal signature that is not directly mirrored in the Eulerian spectra. In particular, the presence of a temporal $1/f$ range does not imply the existence of a corresponding Eulerian $1/k$ range. Indeed, Figure~\ref{fig:taylor_B01B06} shows that, after converting frequency into an effective wavenumber, the temporal $1/f$ interval extends into a low-wavenumber range where the Eulerian spectra do not exhibit an analogous $1/k$ scaling. This difference arises because the Eulerian spectrum describes the spatial distribution of fluctuations within the box at a given time, whereas the probe spectra are built from long trajectories that repeatedly traverse the periodic domain. As a result, the measured temporal signal accumulates contributions from successive crossings of different turbulent regions, rather than representing a single instantaneous spatial cut of the simulation. In this sense, the low-frequency temporal range should not be interpreted as the direct recovery of a spatial low-$k$ power law, but rather as a feature produced by the long-time sampling procedure itself.

This point is related to the superposition principles discussed by \citet{Wang2026b}, who showed that a $1/f$ spectrum can be generated or preserved through different ways of combining signals, including the concatenation of time series. In the solar wind, such superposition is associated with the fact that long-duration spacecraft measurements necessarily include plasma patches originating from different source regions and subsequently mixed during heliospheric transport. Although our numerical setup does not model this heliospheric mixing explicitly, the virtual probe measurements provide a controlled analogue in which long temporal records are built by repeated sampling of a periodic turbulent domain. From this perspective, the difference between the Eulerian spatial spectra and the probe temporal spectra suggests that the measured temporal $1/f$ range should not be regarded simply as the temporal projection of a pre-existing spatial power law. Instead, it appears as a feature shaped by the combined effects of anisotropy, advection, sampling geometry, intrinsic plasma frame dynamics, and the effective concatenation of successive sampled turbulent structures.

In previous observational studies on the applicability of Taylor's hypothesis in the solar wind, single-spacecraft frequency spectra are commonly interpreted as spatial spectra under the assumption that plasma structures are effectively frozen and advected past the observer, so that temporal and spatial scales are related through the sampling velocity \citep{BC2013,Chen_2016}. However, both recent and earlier observational works have emphasized that the validity of this approximation is not uniform across solar wind conditions. On the one hand, \citet{Perez2021} showed, using PSP data, that the recovery of the underlying spatial spectrum is favored when the sampling is sufficiently oblique with respect to the local magnetic field, so that the measured frequency spectral indices can still provide a reliable representation of the plasma frame spectrum. On the other hand, \citet{Perri_2010}, extending previous results by \citet{M1982}, found that the applicability of the Taylor hypothesis becomes only marginal in mixed-stream intervals, where strong shears, compressions, and departures from weak stationarity make the interpretation of time series in purely spatial terms less robust. From this perspective, our numerical results are qualitatively consistent with the idea that the success of the Taylor mapping depends not only on the magnitude of the sampling speed, but also on the sampling geometry relative to the magnetic field, which controls the anisotropic character of the measured fluctuations \citep{Horbury2008,Turner2011}. In particular, the progressively better collapse of the rescaled probe spectra onto the Eulerian spectra at larger $V/V_A$, together with the stronger agreement for nearly perpendicular sampling in Run II, at least in the inertial and dissipation ranges, supports the view that the temporal signal is more reliably interpreted as a sampling of spatial structure when advection dominates over the intrinsic dynamics of the fluctuations. Although the present analysis is based on single-probe sampling, this limitation also underscores the broader importance of multipoint strategies, which can access spatial statistics more directly and reduce the reliance on Taylor assumptions, particularly in anisotropic turbulence (see e.g., \citet{Pecora_2024}).


In addition, the spectral rescaling analysis provides a complementary way to interpret this velocity dependence. This test asks whether spectra measured at different probe velocities can be related to each other through the simple frequency and amplitude transformation expected from Taylor's hypothesis. For Run II, the rescaled spectra collapse more closely onto the reference spectrum as the probe velocity increases, whereas the collapse remains poorer for the slowest probes. This result reinforces the idea that the sampling velocity controls not only the apparent frequency range over which the $1/f$ spectrum is observed, but also the degree to which the measured signal can be interpreted as the advection of an approximately frozen turbulent field. This trend is qualitatively consistent with the slow-flow regime discussed by \citet{Klein_2014}. Using synthetic spacecraft measurements, the authors showed that when the advection speed is not sufficiently large compared with the characteristic Alfv\'enic propagation speed, the frequency in the plasma frame can no longer be neglected in the transformation to the measured frequency, and the resulting spectrum is shifted toward higher frequencies with respect to the Taylor prediction while approximately preserving its spectral slope. Although our setup is different, the behavior of our slowest probes is compatible with the same general interpretation. In particular, the fact that the simple rescaling performs worst for the lowest values of $V/V_A$ suggests that, in that regime, the measured temporal signal is no longer controlled only by advection across an approximately frozen spatial structure. Instead, intrinsic temporal dynamics begin to contribute more significantly to the observed spectrum, preventing a simple collapse under the Taylor-based transformation. From this perspective, the improved agreement obtained for faster probes supports the idea that the dominant effect of changing the probe speed is indeed a frequency shift associated with sampling, whereas the discrepancies at low $V/V_A$ reflect the progressive breakdown of that approximation. It is worth noting that, although we focus on the $1/f$ range whereas \citet{Klein_2014} considered the inertial and kinetic-scale ranges, both analyses point to the same basic effect: changes in the sampling speed modify the observed frequency spectrum. In that sense, our results suggest that the slow-flow argument may remain relevant beyond the inertial and kinetic regimes, extending as well to the low-frequency MHD range explored by our virtual probes. More generally, this interpretation is also in line with previous work showing that relaxing the frozen-flow assumption mainly introduces Doppler-like shifts and frequency broadening, rather than a complete modification of the underlying spectral scaling \citep{Wilczek2012,Everard2021}.

More broadly, the origin and development of the $1/f$ range remain open questions. Although low-frequency $1/f$ spectra are commonly observed in the solar wind, it is still unclear to what extent they reflect source-region processes, nonlinear evolution during solar wind expansion, sampling effects, or the superposition of structures with different correlation properties. In forthcoming works, the virtual probe framework could be used to test superposition scenarios more directly by separating individual box-crossing intervals, estimating their correlation times, and constructing new synthetic records through controlled concatenation or random truncation procedures, in analogy with the methods explored by \citet{Wang2026a}. Such an analysis would make it possible to determine whether the temporal $1/f$ range found here is mainly controlled by the spatial organization of the MHD turbulence, by the sampling geometry and velocity, or by the effective superposition of multiple sampled intervals with distinct characteristic time scales. A complementary direction will be to compare these numerical results with in situ spacecraft measurements, applying similar spectral diagnostics and sampling criteria to solar wind intervals observed under different Alfvénic conditions and at different heliocentric distances.

\section*{Acknowledgements}

The author(s) would like to thank Fouad Sahraoui for fruitful scientific discussions. All authors acknowledge financial support from the following grants: ECOS SUD 2022 $\#$ A22U02 CNRS / CONICET grant. PIP Grant No.~11220200101752, UBACyT Grant No.~20020220300122BA and Redes de Alto Impacto REMATE from Argentina. 
\section*{Data availability}

The data underlying this article were generated from direct numerical simulations. The data are not publicly available due to their large volume, but will be shared by the corresponding author upon reasonable request.


\bibliographystyle{mnras}
\bibliography{main}

@article{Perez2021,
    author = "Perez, Jean C. and Bourouaine, Sofiane and Chen, Christopher H. K. and Raouafi, Nour E.",
    title = "Applicability of Taylor’s hypothesis during Parker Solar Probe perihelia",
    DOI= "10.1051/0004-6361/202039879",
    url= "https://doi.org/10.1051/0004-6361/202039879",
    journal = "A\&A",
    year = 2021,
    volume = 650,
    pages = "A22",
}

@article{Huang2023,
doi = {10.3847/2041-8213/acd7f2},
url = {https://dx.doi.org/10.3847/2041-8213/acd7f2},
year = {2023},
month = {jun},
publisher = {The American Astronomical Society},
volume = {950},
number = {1},
pages = {L8},
author = {Zesen Huang and Nikos Sioulas and Chen Shi and Marco Velli and Trevor Bowen and Nooshin Davis and B. D. G. Chandran and Lorenzo Matteini and Ning Kang and Xiaofei Shi and Jia Huang and Stuart D. Bale and J. C. Kasper and Davin E. Larson and Roberto Livi and P. L. Whittlesey and Ali Rahmati and Kristoff Paulson and M. Stevens and A. W. Case and Thierry Dudok de Wit and David M. Malaspina and J. W. Bonnell and Keith Goetz and Peter R. Harvey and Robert J. MacDowall},
title = {New Observations of Solar Wind 1/f Turbulence Spectrum from Parker Solar Probe},
journal = {The Astrophysical Journal Letters}
}

@article{M1986,
  title = {Low-Frequency $\frac{1}{f}$ Noise in the Interplanetary Magnetic Field},
  author = {Matthaeus, W. H. and Goldstein, M. L.},
  journal = {Phys. Rev. Lett.},
  volume = {57},
  issue = {4},
  pages = {495--498},
  numpages = {0},
  year = {1986},
  month = {Jul},
  publisher = {American Physical Society},
  doi = {10.1103/PhysRevLett.57.495},
  url = {https://link.aps.org/doi/10.1103/PhysRevLett.57.495}
}

@article{C2018, title={Parametric instability, inverse cascade and the $1/f$ range of solar-wind turbulence}, volume={84}, DOI={10.1017/S0022377818000016}, number={1}, journal={Journal of Plasma Physics}, author={Chandran, Benjamin D. G.}, year={2018}, pages={905840106}}

@article{Velli1989,
  title = {Turbulent cascade of incompressible unidirectional Alfv\'en waves in the interplanetary medium},
  author = {Velli, Marco and Grappin, Roland and Mangeney, Andr\'e},
  journal = {Phys. Rev. Lett.},
  volume = {63},
  issue = {17},
  pages = {1807--1810},
  numpages = {0},
  year = {1989},
  month = {Oct},
  publisher = {American Physical Society},
  doi = {10.1103/PhysRevLett.63.1807},
  url = {https://link.aps.org/doi/10.1103/PhysRevLett.63.1807}
}

@article{Verdini2012,
doi = {10.1088/2041-8205/750/2/L33},
url = {https://dx.doi.org/10.1088/2041-8205/750/2/L33},
year = {2012},
month = {apr},
publisher = {The American Astronomical Society},
volume = {750},
number = {2},
pages = {L33},
author = {Verdini, Andrea and Grappin, Roland and Pinto, Rui and Velli, Marco},
title = {ON THE ORIGIN OF THE 1/f SPECTRUM IN THE SOLAR WIND MAGNETIC FIELD},
journal = {The Astrophysical Journal Letters}

}

@article{Matteini2018,
doi = {10.3847/2041-8213/aaf573},
url = {https://dx.doi.org/10.3847/2041-8213/aaf573},
year = {2018},
month = {dec},
publisher = {The American Astronomical Society},
volume = {869},
number = {2},
pages = {L32},
author = {Matteini, L. and Stansby, D. and Horbury, T. S. and Chen, C. H. K.},
title = {On the 1/f Spectrum in the Solar Wind and Its Connection with Magnetic Compressibility},
journal = {The Astrophysical Journal Letters}
}

@ARTICLE{BC2013,
author = {{Bruno}, Roberto and {Carbone}, Vincenzo},
title = "{The Solar Wind as a Turbulence Laboratory}",
journal = {Living Reviews in Solar Physics},
year = 2013,
month = dec,
volume = {10},
number = {1},
eid = {2},
pages = {2},
doi = {10.12942/lrsp-2013-2},
adsurl = {https://ui.adsabs.harvard.edu/abs/2013LRSP...10....2B}
}

@article{D2007,
  title = {Low-frequency $1∕f$ fluctuations in hydrodynamic and magnetohydrodynamic turbulence},
  author = {Dmitruk, Pablo and Matthaeus, W. H.},
  journal = {Phys. Rev. E},
  volume = {76},
  issue = {3},
  pages = {036305},
  numpages = {8},
  year = {2007},
  month = {Sep},
  publisher = {American Physical Society},
  doi = {10.1103/PhysRevE.76.036305},
  url = {https://link.aps.org/doi/10.1103/PhysRevE.76.036305}
}

@article{Chen_2020,
doi = {10.3847/1538-4365/ab60a3},
url = {https://dx.doi.org/10.3847/1538-4365/ab60a3},
year = {2020},
month = {feb},
publisher = {The American Astronomical Society},
volume = {246},
number = {2},
pages = {53},
author = {Chen, C. H. K. and Bale, S. D. and Bonnell, J. W. and Borovikov, D. and Bowen, T. A. and Burgess, D. and Case, A. W. and Chandran, B. D. G. and de Wit, T. Dudok and Goetz, K. and Harvey, P. R. and Kasper, J. C. and Klein, K. G. and Korreck, K. E. and Larson, D. and Livi, R. and MacDowall, R. J. and Malaspina, D. M. and Mallet, A. and McManus, M. D. and Moncuquet, M. and Pulupa, M. and Stevens, M. L. and Whittlesey, P.},
title = {The Evolution and Role of Solar Wind Turbulence in the Inner Heliosphere},
journal = {The Astrophysical Journal Supplement Series}
}

@article{Bro2023,
    author = {Brodiano, M. and Dmitruk, P. and Andrés, N.},
    title = {A statistical study of the compressible energy cascade rate in solar wind turbulence: Parker solar probe observations},
    journal = {Physics of Plasmas},
    volume = {30},
    number = {3},
    pages = {032903},
    year = {2023},
    month = {03},
    issn = {1070-664X},
    doi = {10.1063/5.0109379},
    url = {https://doi.org/10.1063/5.0109379},
    eprint = {https://pubs.aip.org/aip/pop/article-pdf/doi/10.1063/5.0109379/16796315/032903\_1\_online.pdf},
}

@article{Dorseth2024b,
author = {{Dorseth, Mason} and {Perez, Jean C.} and {Bourouaine, Sofiane} and {Palacios, Juan C.} and {Raouafi, Nour E.}},
title = {The low-frequency power spectrum of slow solar wind turbulence},
DOI= "10.1051/0004-6361/202449869",
url= "https://doi.org/10.1051/0004-6361/202449869",
journal = {A$\&$A},
year = 2024,
volume = 689,
pages = "A117",
}

@article{ZhouEA2004,
title={Magnetohydrodynamic turbulence and time scales in astrophysical and space plasmas},
volume={76},
DOI={10.1103/RevModPhys.76.1015},
journal={Reviews of Modern Physics},
author={Zhou, Y. and Matthaeus, W.H. and Dmitruk, P.},
year={2004},
pages={1015}}

@article{M1982,
author = {Matthaeus, William H. and Goldstein, Melvyn L.},
title = {Stationarity of magnetohydrodynamic fluctuations in the solar wind},
journal = {Journal of Geophysical Research: Space Physics},
volume = {87},
number = {A12},
pages = {10347-10354},
doi = {https://doi.org/10.1029/JA087iA12p10347},
url = {https://agupubs.onlinelibrary.wiley.com/doi/abs/10.1029/JA087iA12p10347},
eprint = {https://agupubs.onlinelibrary.wiley.com/doi/pdf/10.1029/JA087iA12p10347},
year = {1982}
}

@article{D2024,
	author = {{Dorseth, Mason} and {Perez, Jean C.} and {Bourouaine, Sofiane} and {Palacios, Juan C.} and {Raouafi, Nour E.}},
	title = {The low-frequency power spectrum of slow solar wind turbulence},
	DOI= "10.1051/0004-6361/202449869",
	url= "https://doi.org/10.1051/0004-6361/202449869",
	journal = {A$\&$A},
	year = 2024,
	volume = 689,
	pages = "A117",
}

@misc{Bro2026,
      title={An Intermittent Model for the $1/f$ Spectrum in the Pristine Solar Wind}, 
      author={Maia Brodiano and Fouad Sahraoui and Davide Manzini and Lina Z. Hadid and Facundo Pugliese and Pablo Dmitruk and Nahuel Andrés},
      year={2025},
      eprint={2506.04366},
      archivePrefix={arXiv},
      primaryClass={astro-ph.SR},
      url={https://arxiv.org/abs/2506.04366}, 
}

@article{Go2005b,
author = {Gómez, D. O. and Mininni, P. D. and Dmitruk, P. A.},
year = {2006},
month = {03},
pages = {123},
title = {Parallel Simulations in Turbulent MHD},
volume = {2005},
journal = {Physica Scripta},
}

@article{Mi2011,
    author = {Mininni, P. D. and Rosenberg, D. and Reddy, R. and Pouquet, A.},
	title = {A hybrid MPI-OpenMP scheme for scalable parallel pseudospectral computations for fluid turbulence},
	journal = {Parallel Computing},
	volume = {37},
	number = {6-7},
	pages = {16-326},
	year = {2011},
	
}

@article{Davis2023,
doi = {10.3847/1538-4357/acd177},
url = {https://dx.doi.org/10.3847/1538-4357/acd177},
year = {2023},
month = {jun},
publisher = {The American Astronomical Society},
volume = {950},
number = {2},
pages = {154},
author = {Davis, Nooshin and Chandran, B. D. G. and Bowen, T. A. and Badman, S. T. and de Wit, T. Dudok and Chen, C. H. K. and Bale, S. D. and Huang, Zesen and Sioulas, Nikos and Velli, Marco},
title = {The Evolution of the 1/f Range within a Single Fast-solar-wind Stream between 17.4 and 45.7 Solar Radii},
journal = {The Astrophysical Journal}
}

@article{Sa2023,
doi = {10.3847/2041-8213/acc531},
url = {https://doi.org/10.3847/2041-8213/acc531},
year = {2023},
month = {apr},
publisher = {The American Astronomical Society},
volume = {946},
number = {2},
pages = {L44},
author = {Šafránková, Jana and Němeček, Zdeněk and Němec, František and Verscharen, Daniel and Horbury, Timothy S. and Bale, Stuart D. and Přech, Lubomír},
title = {Evolution of Magnetic Field Fluctuations and Their Spectral Properties within the Heliosphere: Statistical Approach},
journal = {The Astrophysical Journal Letters}
}

@article{Perri_2010,
doi = {10.1088/0004-637X/714/1/937},
url = {https://doi.org/10.1088/0004-637X/714/1/937},
year = {2010},
month = {apr},
publisher = {The American Astronomical Society},
volume = {714},
number = {1},
pages = {937},
author = {Perri, S. and Balogh, A.},
title = {STATIONARITY IN SOLAR WIND FLOWS},
journal = {The Astrophysical Journal}
}

@ARTICLE{Taylor1938,
       author = {{Taylor}, G.~I.},
        title = "{The Spectrum of Turbulence}",
      journal = {Proceedings of the Royal Society of London Series A},
         year = 1938,
        month = feb,
       volume = {164},
       number = {919},
        pages = {476-490},
          doi = {10.1098/rspa.1938.0032},
       adsurl = {https://ui.adsabs.harvard.edu/abs/1938RSPSA.164..476T}
}

@article{Chen_2016, title={Recent progress in astrophysical plasma turbulence from solar wind observations}, volume={82}, DOI={10.1017/S0022377816001124}, number={6}, journal={Journal of Plasma Physics}, author={Chen, C. H. K.}, year={2016}, pages={535820602}}

@article{Pecora_2024,
doi = {10.3847/2041-8213/ad5fff},
url = {https://doi.org/10.3847/2041-8213/ad5fff},
year = {2024},
month = {jul},
publisher = {The American Astronomical Society},
volume = {970},
number = {2},
pages = {L36},
author = {Pecora, Francesco and Pucci, Francesco and Malara, Francesco and Klein, Kristopher G. and Marcucci, Maria Federica and Retinò, Alessandro and Matthaeus, William},
title = {Evaluation of Scale-dependent Kurtosis with HelioSwarm},
journal = {The Astrophysical Journal Letters}
}

@article{Klein_2014,
doi = {10.1088/2041-8205/790/2/L20},
url = {https://doi.org/10.1088/2041-8205/790/2/L20},
year = {2014},
month = {jul},
publisher = {The American Astronomical Society},
volume = {790},
number = {2},
pages = {L20},
author = {Klein, K. G. and Howes, G. G. and TenBarge, J. M.},
title = {THE VIOLATION OF THE TAYLOR HYPOTHESIS IN MEASUREMENTS OF SOLAR WIND TURBULENCE},
journal = {The Astrophysical Journal Letters}
}

@article{Wilczek2012,
  title = {Wave-number--frequency spectrum for turbulence from a random sweeping hypothesis with mean flow},
  author = {Wilczek, M. and Narita, Y.},
  journal = {Phys. Rev. E},
  volume = {86},
  issue = {6},
  pages = {066308},
  numpages = {8},
  year = {2012},
  month = {Dec},
  publisher = {American Physical Society},
  doi = {10.1103/PhysRevE.86.066308},
  url = {https://link.aps.org/doi/10.1103/PhysRevE.86.066308}
}

@article{Everard2021,
author = {Everard, K. A. and Katul, G. G. and Lawrence, G. A. and Christen, A. and Parlange, M. B.},
title = {Sweeping Effects Modify Taylor’s Frozen Turbulence Hypothesis for Scalars in the Roughness Sublayer},
journal = {Geophysical Research Letters},
volume = {48},
number = {22},
pages = {e2021GL093746},
doi = {https://doi.org/10.1029/2021GL093746},
url = {https://agupubs.onlinelibrary.wiley.com/doi/abs/10.1029/2021GL093746},
eprint = {https://agupubs.onlinelibrary.wiley.com/doi/pdf/10.1029/2021GL093746},
note = {e2021GL093746 2021GL093746},
year = {2021}
}

@article{Turner2011,
  title = {Nonaxisymmetric Anisotropy of Solar Wind Turbulence},
  author = {Turner, A. J. and Gogoberidze, G. and Chapman, S. C. and Hnat, B. and M\"uller, W.-C.},
  journal = {Phys. Rev. Lett.},
  volume = {107},
  issue = {9},
  pages = {095002},
  numpages = {5},
  year = {2011},
  month = {Aug},
  publisher = {American Physical Society},
  doi = {10.1103/PhysRevLett.107.095002},
  url = {https://link.aps.org/doi/10.1103/PhysRevLett.107.095002}
}

@article{Horbury2008,
  title = {Anisotropic Scaling of Magnetohydrodynamic Turbulence},
  author = {Horbury, Timothy S. and Forman, Miriam and Oughton, Sean},
  journal = {Phys. Rev. Lett.},
  volume = {101},
  issue = {17},
  pages = {175005},
  numpages = {4},
  year = {2008},
  month = {Oct},
  publisher = {American Physical Society},
  doi = {10.1103/PhysRevLett.101.175005},
  url = {https://link.aps.org/doi/10.1103/PhysRevLett.101.175005}
}

@article{Osman2011,
author = {Osman, Kareem and Wan, Minping and Matthaeus, W. and Breech, B. and Oughton, S.},
year = {2011},
month = {10},
pages = {75},
title = {DIRECTIONAL ALIGNMENT AND NON-GAUSSIAN STATISTICS IN SOLAR WIND TURBULENCE},
volume = {741},
journal = {The Astrophysical Journal},
doi = {10.1088/0004-637X/741/2/75}
}

@article{Pecora_2023,
doi = {10.3847/2041-8213/acbb03},
url = {https://doi.org/10.3847/2041-8213/acbb03},
year = {2023},
month = {mar},
publisher = {The American Astronomical Society},
volume = {945},
number = {2},
pages = {L20},
author = {Pecora, Francesco and Servidio, Sergio and Primavera, Leonardo and Greco, Antonella and Yang, Yan and Matthaeus, William H.},
title = {Multipoint Turbulence Analysis with HelioSwarm},
journal = {The Astrophysical Journal Letters}
}

@article{Matthaeus_2007,
doi = {10.1086/513075},
url = {https://doi.org/10.1086/513075},
year = {2007},
month = {feb},
publisher = {},
volume = {657},
number = {2},
pages = {L121},
author = {Matthaeus, W. H. and Breech, B. and Dmitruk, P. and Bemporad, A. and Poletto, G. and Velli, M. and Romoli, M.},
title = {Density and Magnetic Field Signatures of Interplanetary 1/f Noise},
journal = {The Astrophysical Journal}
}

@article{Wang2026b,
    author = {Wang, Jiaming and Pecora, Francesco and Chhiber, Rohit and Pradata, Rayta A and Adhikari, Subash and Matthaeus, William H},
    title = {1/f noise in synthetic and solar wind data: superposition principles},
    journal = {Monthly Notices of the Royal Astronomical Society},
    volume = {548},
    number = {4},
    pages = {stag722},
    year = {2026},
    month = {06},
    issn = {0035-8711},
    doi = {10.1093/mnras/stag722},
    url = {https://doi.org/10.1093/mnras/stag722},
    eprint = {https://academic.oup.com/mnras/article-pdf/548/4/stag722/68106543/stag722.pdf},
}

@article{Wang2026a,
author = {Wang, Jiaming and Pecora, Francesco and Chhiber, Rohit and Roy, Sohom and Matthaeus, W.},
year = {2026},
month = {01},
pages = {},
title = {Interplanetary magnetic correlation and low-frequency spectrum over many solar rotations},
volume = {123},
journal = {Proceedings of the National Academy of Sciences},
doi = {10.1073/pnas.2519811122}
}

@article{Wicks2013,
  author = {Wicks, R. T. and Roberts, D. A. and Mallet, A. and Schekochihin, A. A. and Horbury, T. S. and Chen, C. H. K.},
  title = {Correlations at Large Scales and the Onset of Turbulence in the Fast Solar Wind},
  journal = {The Astrophysical Journal},
  volume = {778},
  number = {2},
  pages = {177},
  year = {2013},
  doi = {10.1088/0004-637X/778/2/177}
}

@article{Huang2025,
  author = {Huang, Zesen and Velli, Marco and Chandran, B. D. G. and Shi, Chen and Ding, Yuliang and Matteini, Lorenzo and Choi, Kyung-Eun},
  title = {Two Types of 1/f Range in Solar Wind Turbulence},
  journal = {The Astrophysical Journal Letters},
  volume = {990},
  pages = {L34},
  year = {2025},
  doi = {10.3847/2041-8213/adfa13}
}

@article{Ba1982,
author = {Bavassano, B. and Dobrowolny, M. and Mariani, F. and Ness, N. F.},
title = {Radial evolution of power spectra of interplanetary Alfvénic turbulence},
journal = {Journal of Geophysical Research: Space Physics},
volume = {87},
number = {A5},
pages = {3617-3622},
doi = {https://doi.org/10.1029/JA087iA05p03617},
url = {https://agupubs.onlinelibrary.wiley.com/doi/abs/10.1029/JA087iA05p03617},
eprint = {https://agupubs.onlinelibrary.wiley.com/doi/pdf/10.1029/JA087iA05p03617},
year = {1982}
}

@ARTICLE{Burlaga_1984,
       author = {{Burlaga}, L.~F. and {Goldstein}, M.~L.},
        title = "{Radial variations of large-scale magnetohydrodynamic fluctuations in the solar wind}",
      journal = {\jgr},
         year = 1984,
        month = aug,
       volume = {89},
       number = {A8},
        pages = {6813-6818},
          doi = {10.1029/JA089iA08p06813},
       adsurl = {https://ui.adsabs.harvard.edu/abs/1984JGR....89.6813B}
}

@article{Magyar_2022,
doi = {10.3847/1538-4357/ac8b81},
url = {https://doi.org/10.3847/1538-4357/ac8b81},
year = {2022},
month = {oct},
publisher = {The American Astronomical Society},
volume = {938},
number = {2},
pages = {98},
author = {Magyar, Norbert and Doorsselaere, Tom Van},
title = {Phase Mixing and the 1/f Spectrum in the Solar Wind},
journal = {The Astrophysical Journal}
}

@article{Pradata_2025,
doi = {10.3847/2041-8213/adc9b2},
url = {https://doi.org/10.3847/2041-8213/adc9b2},
year = {2025},
month = {apr},
publisher = {The American Astronomical Society},
volume = {984},
number = {1},
pages = {L23},
author = {Pradata, Rayta A. and Roy, Sohom and Matthaeus, William H. and Wang, Jiaming and Chhiber, Rohit and Pecora, Francesco and Yang, Yan},
title = {Observations of 1/f Noise at Mercury’s Solar Wind Using MESSENGER Data},
journal = {The Astrophysical Journal Letters}
}

@article{Chuychai2014,
author = {Chuychai, P. and Weygand, J. M. and Matthaeus, W. H. and Dasso, S. and Smith, C. W. and Kivelson, M. G.},
title = {Technique for measuring and correcting the Taylor microscale},
journal = {Journal of Geophysical Research: Space Physics},
volume = {119},
number = {6},
pages = {4256-4265},
doi = {https://doi.org/10.1002/2013JA019641},
url = {https://agupubs.onlinelibrary.wiley.com/doi/abs/10.1002/2013JA019641},
eprint = {https://agupubs.onlinelibrary.wiley.com/doi/pdf/10.1002/2013JA019641},
year = {2014}
}

@article{Ghanbari_2023,
doi = {10.3847/1538-4357/acabc4},
url = {https://doi.org/10.3847/1538-4357/acabc4},
year = {2023},
month = {jan},
publisher = {The American Astronomical Society},
volume = {943},
number = {2},
pages = {87},
author = {Ghanbari, Keyvan and Florinski, Vladimir},
title = {Simulation of Solar Wind Turbulence near Corotating Interaction Regions: Superposed Epoch Analysis of Simulations and Observations},
journal = {The Astrophysical Journal}
}

@article{Bro2021,
doi = {10.3847/1538-4357/ac2834},
url = {https://doi.org/10.3847/1538-4357/ac2834},
year = {2021},
month = {dec},
publisher = {The American Astronomical Society},
volume = {922},
number = {2},
pages = {240},
author = {Brodiano, M. and Andrés, N. and Dmitruk, P.},
title = {Spatiotemporal Analysis of Waves in Compressively Driven Magnetohydrodynamics Turbulence},
journal = {The Astrophysical Journal}
}

@article{Pugliese2023,
doi = {10.3847/1538-4357/ad055b},
url = {https://doi.org/10.3847/1538-4357/ad055b},
year = {2023},
month = {dec},
publisher = {The American Astronomical Society},
volume = {959},
number = {1},
pages = {28},
author = {Pugliese, F. and Brodiano, M. and Andrés, N. and Dmitruk, P.},
title = {Energization of Charged Test Particles in Magnetohydrodynamic Fields: Waves versus Turbulence Picture},
journal = {The Astrophysical Journal}
}

\label{lastpage}
\end{document}